\documentclass[twoside,12pt]{article}

\usepackage{graphicx}

\begin{document}

\title{\vspace{1cm} The Cosmological QCD Phase Transition Revisited}

\author{Tillmann Boeckel, Simon Schettler, J\"urgen Schaffner-Bielich\\
\\
Institute for Theoretical Physics, Heidelberg University,\\
Philosophenweg 16, 69120 Heidelberg, Germany}

\maketitle

\begin{abstract} 
  The QCD phase diagram might exhibit a first order phase transition
  for large baryochemical potentials. We explore the cosmological
  implications of such a QCD phase transition in the early
  universe. We propose that the large baryon-asymmetry is diluted by a
  little inflation where the universe is trapped in a false vacuum
  state of QCD.  The little inflation is stopped by bubble nucleation
  which leads to primordial production of the seeds of extragalactic
  magnetic fields, primordial black holes and gravitational waves. In
  addition the power spectrum of cold dark matter can be affected up
  to mass scales of $10^9 M_\odot$. The imprints of the cosmological
  QCD phase transition on the gravitational wave background can be
  explored with the future gravitational wave detectors LISA and BBO
  and with pulsar timing.
\end{abstract}


\section{Introduction}


The history of the early universe left an imprint on the presently
observed cosmos. According to the Friedmann equations the temperature
increases inversely proportional with scale parameter $a$ so that the
early universe passes through the big bang nucleosynthesis (BBN) at $t=
1$s to 3 minutes (corresponding to $T=0.1$ to 1 MeV), the QCD phase
transition at $t\approx 10^{-5}$s ($T\approx 150$ MeV) and the
electroweak phase transition at $t\approx 10^{-10}$s ($T\approx 100$
GeV). BBN and the cosmological electroweak phase transition have
received considerable attention in the last years by studying the
production of light elements and baryogenesis. On the other hand the
cosmological QCD phase transition does not seem to be associated with
a key observable in todays universe. In this contribution we revisit
the cosmological QCD phase transition and discuss some new
cosmological signals in view of the little inflation scenario proposed
by us \cite{Boeckel:2009ej,Schettler:2010dp}. The basic ingredients for a
short inflationary period are an Affleck-Dine-type baryogenesis, in
order to achieve a large initial baryon-to-photon ratio, and a
metastable vacuum due to the nonvanishing vacuum expectation values of
QCD at high net baryon densities. 

The implications and possible signals of such a scenario are
surprisingly rich and involve large-scale structure formation up to
dwarf galaxy scales, the production of dark matter (WIMPs) and mini
black holes, provide possibly the seeds of the cosmological magnetic
fields and leave an imprint on the gravitational wave background. The
latter signal is particular interesting, as it can be probed with the
gravitational wave detectors LISA and BBO and with pulsar timing. We
point out that the QCD phase transition without inflation has been
widely discussed in the literature before although assuming a first
order phase transition at nearly vanishing baryon densities (see e.g.\
\cite{Hogan:1983zz,Hogan:1984hx,Witten:1984rs,Kamionkowski:1993fg,Cheng:1994yr,Jedamzik:1996mr,Sigl:1996dm,Schmid:1998mx,Huber:2008hg,Caprini:2009pr,Caprini:2010xv}
and \cite{Schwarz:2003du} for a review) so that these results can be
transferred to the little inflation scenario presented in the
following.  The concept of a mini-inflation (or tepid inflation) at
the QCD phase transition has been also introduced earlier by K\"ampfer
et al.\
\cite{Kampfer,Boiko:1990,Jenkovszky:1990ex,Kampfer:2000gx}
and for a substantially stronger inflationary period by Borghini et
al.\ \cite{Borghini:2000yp} which in both cases will dilute an initial
high net baryon density to the presently observed small value of the
baryon-to-photon ratio.


\section{A Little Inflation at the QCD Phase Transition}


In the standard cosmology the early universe passes through the QCD
phase diagram at small baryon density and high temperature, a region
which is presently probed with ultrarelativistic heavy-ion collisions
at BNL's RHIC and CERN's LHC. The other part of the QCD phase diagram,
high baryon densities and moderate to low temperatures, is the realm
of core-collapse supernovae and neutron stars and will be investigated
in more detail in terrestrial experiments with the upcoming Compressed
Baryonic Experiment CBM at GSI's Facility for Antiproton and Ion
Research FAIR.

From the analysis of the microwave background radiation and big bang
nucleosynthesis one deduces a baryon-to-photon ratio of $n_B/s \sim
n_B/n_\gamma \sim \mu/T \sim 10^{-9}$, a ratio which is conserved
after baryogenesis and the last first order non-equilibrium phase
transition in the early universe as entropy is conserved during the
standard cosmological evolution. Hence, the early universe evolves
along $\mu/T \sim 10^{-9}\sim 0$ in the QCD phase diagram. As lattice
data indicates a crossover transition at vanishing baryochemical
potential \cite{Aoki:2006we}, nothing spectacular is happening and no
particular strong cosmological signals are expected from the QCD phase
transition.

The Friedmann equation for a radiation dominated universe reads 
\[ H^2 = \frac{8\pi G}{3} \rho \sim g(T) \frac{T^4}{M_p^2} \] 
where $g(T)$ is the effective number of relativistic degrees of
freedom at the temperature $T$. The Hubble time is related to the true
time by $t=3t_H$ for a radiation dominated universe and is given by
\[ t_H = \frac{1}{H} \sim  g^{-1/2} \frac{M_P}{T^2} 
\Longrightarrow \frac{t}{\mbox{1 sec}} \sim \left(\frac{\mbox{1 MeV}}{T} \right)^2
\]
So for the QCD phase transition at $T=T_c\sim 150$~MeV one arrives at a
time of about $10^{-5}$~s after the big bang.

We explore in the following the scenario that the early universe
passed through a first order QCD phase transition which is presently
advocated for large baryochemical potentials. First, we address the
question whether this is possible or not with the present cosmological
data. Second, we delineate possible cosmological signals of a first
order QCD phase transition which could be observable in the future.

For a first order phase transition there exists a false metastable
vacuum, in which the universe could be trapped during cooldown. As the
total energy density is then determined by the constant potential of
this false vacuum state, the universe evolves according to de Sitter
solution. One can easily derive from the Friedmann equation that a
constant energy density implies an exponential expansion rate
\[ H=\dot a/a \sim M_p^{-1}\rho_{\rm v}^{1/2} = H_{\rm v} = {\rm
  const.}  \to a \sim \exp (H_{\rm v}\cdot t) \] 
so that the universe goes through another inflationary epoch. During
this time the universe supercools and the density decreases
exponentially while the ratio $\mu/T$ is preserved.  At the end of
inflation, the universe falls into the true vacuum state which
releases latent heat. The universe gets reheated with a final
temperature similar to the one at the start of inflation. The
parameters are such that the temperature and baryochemical potential
are now below the phase transition line.  The corresponding increase
in photon density results in a reduced baryon-to-photon ratio.  For
our purposes just a few e-folds are enough to dilute the baryon to
photon ratio to the presently observed value of about $10^{-9}$. Then
for a gas of massless quarks, the baryon-to-photon ratio before and
after inflation are related by:
\[ \left(\frac{\mu}{T}\right)_f \approx \left(\frac{a_i}{a_f}\right)^3
\left( \frac{\mu}{T}\right)_i \] 
as the baryon density is diluted by the scale factor cubed. Hence an
initial ratio of $\left(\mu/T\right)_i \sim \cal{O}$(1) can be reached
for just $N = \ln \left(a_f/a_i\right) \sim \ln (10^3) \sim 7$ e-folds
(for comparison standard inflation at the GUT scale needs $N\sim 50$
e-folds). For such high values of the baryon-to-photon ratio a first
order QCD phase transition is presently discussed in the literature in
connection with chiral symmetry restoration. A first order QCD phase
transition in the early universe could have also been triggered by a
large lepton asymmetry as pointed out recently by Schwarz and Stuke
\cite{Schwarz:2009ii} (for a popular account on the ``bubbling
universe'' at the QCD phase transition see \cite{NewScientist:2010}).
We note in passing that the high baryon asymmetry in our scenario
presumably also implies a correspondingly high lepton asymmetry. A
first-order phase transition in QCD can be modeled within the linear
$\sigma$ model \cite{Pisarski:1983ms} with the possibility of a quench
at finite chemical potentials \cite{Scavenius:1999zc}. The quark
condensate or here the expectation value of the $\sigma$-field serves
as an order parameter. Actually, the QCD phase transition might be
more complex owing to its nontrivial vacuum structure. There is a
second order parameter, the Polyakov loop, which is related to the
gauge sector of the theory and its deconfinement phase transition. The
trace anomaly of QCD can be related to the vacuum expectation value of
the gluon condensate and its scaling properties. So it might well be
that there is a second scalar field, a dilaton field, which has to be
taken into account in an effective description of the phase
transition. Interestingly, such an effective model with two scalar
fields opens the door for a hybrid inflation scenario in cosmology
within the standard model.

A little inflation in the QCD phase diagram starts with a high initial
value of baryon-to-photon ratio which for massless particles is
proportional to the ratio of baryochemical potential to temperature so
that $\mu/T\sim 1$. Such a large value can be inherited from some
earlier nonequilibrium processes producing a net baryon number as in
Affleck-Dine baryogenesis \cite{Affleck:1984fy}. Then the conditions
in the early universe are such that the first order phase transition
line at large baryochemical potentials of QCD is hit and the universe
is trapped in false vacuum state. The constant and nonvanishing vacuum
energy density leads to an inflationary expansion associated with an
exponential supercooling and dilution of matter. Note that during this
evolution the universe is not in an equilibrated state while both the
temperature and the baryochemical potential drop exponentially with
$\mu/T = {\rm const.}$. At low temperatures the barrier to the true
vacuum state becomes so low that the universe rolls down to the true
vacuum state. The released latent heat is converted to particle
production and eventually reheats the universe to temperatures of
$T\sim T_c$. During reheating the baryon number is conserved so that
the net baryon density is still substantially reduced compared to the
initial state before inflation, so that $\mu/T\sim 10^{-9}$, the value
observed today. Afterwards the standard cosmological evolution follows
proceeding to big bang nucleosynthesis.

The evolution of energy densities during the little inflation period
can be discussed in quite general terms. The energy density of
nonrelativistic dark matter falls off as $a^{-3}$ always. The total
energy density, however, is determined by the one of the QCD vacuum
once the universe is trapped in the false high-temperature QCD vacuum
state and inflation starts.  The radiation energy density falls off
initially as $a^{-4}$ until the end of inflation. Then energy is
gained from the phase transition from the false vacuum state to the
true one which is about the energy scale of the QCD vacuum. This
energy is transferred to heat and the production of relativistic
particles which thereby increases the radiation energy density
correspondingly. The baryon density and dark matter density to entropy
ratio will be diluted at the end of inflation due to the production of
entropy. By construction the maximum length of inflation is about the
one given by the initial value of the baryon to entropy ratio which
translates to a ratio of the initial and final value of the scale
parameter of about $a_f/a_i = 10^3$.


\section{Cosmological implications of the QCD phase transition}


A prominent candidate for cold dark matter are weakly interacting
massive particles (WIMPs), which freeze-out while being
non-relativistic with an annihilation cross section similar to the one
expected from weak interactions between e.g.\ SUSY particles. The
present day energy density for WIMPs is just given by the annihilation
cross section with logarithmic corrections from the mass of the WIMP
as $\Omega_{CDM} \sim \sigma_{\rm weak}/\sigma_{\rm ann.}$. If the
number density of dark matter is diluted by a factor $(a_f/a_i)^3 \sim
10^9$ due to the little inflation, the annihilation cross section has
to be reduced accordingly so that the present day abundance is matched
despite the dilution factor. Then the production cross section is
reduced by a similar factor. The observation of the dark matter WIMP
at the LHC would then not be possible in the little inflation
scenario.

The nonstandard evolution of the energy densities will leave an
imprint on the power spectrum of dark matter. The dark matter mass
within the horizon at the critical temperature of QCD of $T_c\approx
150$ MeV is just $10^{-9} M_\odot$, too small a mass scale to have any
cosmological consequences. However, the mass scale is boosted by the
little inflation by at least a factor $(a_f/a_i)^3\approx 10^9$ so
that mass scales of up to $1 M_\odot$ are affected.  There is an
additional effect for modes $k_{ph}<H$ at the beginning of inflation
as there are two scales involved during inflation: one is the Hubble
parameter $H \propto \rho^{1/2}_{\rm v} \sim \mbox{const.}$ and the
other one is 
\[ \dot{H} = -4\pi G(\rho + p) = -4\pi G (\rho_{dm} + 4\rho_r/3)
\propto (a_i/a)^q \]
where $q=3\dots 4$ depending on whether the matter or radiation energy
density is the dominant contribution. Note, that the total energy
density is given by the constant vacuum energy density which, however,
drops out for $\dot{H}$. For standard inflation only the former one,
the Hubble horizon, is important but for a small period of inflation
also the latter scale leaves an observable effect on the power
spectrum.  Correspondingly, there are three different spectral regimes
to be considered:
\begin{itemize}
\item $(k_{ph}/H)_i > a_f/a_i$: always subhubble
\item $a_f/a_i > (k_{ph}/H)_i > (a_i/a_f)^{q/2}$: intermediate
\item $(k_{ph}/H)_i < (a_i/a_f)^{q/2}$: unaffected
\end{itemize}
Hence, the highest mass scale affected can be as large as $M_{max}
\sim 10^{-9} M_\odot (a_f/a_i)^{3q/2} \sim (10^{4.5} - 10^9) M_\odot$
depending on the value of $q$. For interesting inflation lengths the
value of $q$ will be closer to 3 than to 4, because dark matter will
then be more abundant than radiation for most of the inflationary
phase, resulting in a maximum mass of $\sim 10^6 M_\odot$. These
mass scales are of cosmological interest as it reaches the mass scales
of globular clusters and dwarf galaxies. The power spectrum will be
reduced, so that it would be interesting to study its impact on the
cuspy core density distribution of dark matter in small galaxies and
the large number of halo structures seen in standard structure
formation.

The nonequilibrium process ending inflation is likely to produce
bubbles by nucleation or by spinodal decomposition. The fluctuations
in density together with a softening of the equation of state at the
phase transition can create primordial black holes by collapsing
bubbles \cite{Jedamzik:1996mr,Kapusta:2007dn}. The maximum mass of the
black hole is given be the total enclosed energy density at the time
of the QCD phase transition, i.e.\ after inflation, which is about
$M_{bh} \sim M_{hubble} \sim 1 M_\odot$.

The nonequilibrium phase transition and the formation of bubbles will
generate perturbations in the metric. It turns out that the tensor
perturbations are directly related to the QCD trace anomaly.  The
equation of motion for the tensor perturbation amplitude $v_k= a\cdot
h_k$ in Fourier space (gauge invariant) is given in a gauge invariant
way by:
\[ v''_k(\eta) + \left(k^2 - \frac{a''}{a}\right) v_k(\eta) = 0 
\qquad \mbox{ with } \qquad
\frac{a''}{a} = \frac{4\pi G a^2}{3} \left(\rho-3p\right) \] 
Hence, only the trace of the energy-momentum tensor is responsible for
generating tensor perturbation, i.e.\ gravitational waves from the
phase transition. We use several parameterizations of lattice gauge
calculations as input taken from \cite{Bazavov:2009zn,Borsanyi:2010cj}
and compare with the simple bag model.

The energy density in the gravitational wave background is usually
expressed for a given wavenumber $k$ and reads $\Omega_g(k) =
\frac{1}{\rho_c} \frac{d\rho_g}{d\ln k}$. After horizon entry, the
mode $h_k$ is damped by the scale factor as $1/a$.  Using entropy
conservation and the Friedmann equations one can show that the energy
density will exhibit a steplike change due to the different
relativistic degrees of freedom before and after the phase transition
as
\[ \frac{\Omega_g(\nu \gg \nu^*)}{\Omega_g(\nu \ll \nu^*)} =
\left(\frac{g_f}{g_i}\right)^{1/3} \sim 0.7 \]
which has been shown by Schwarz \cite{Schwarz:1997gv}. The
characteristic frequency for the QCD phase transition is redshifted
today by
\[ \nu_{\rm peak} \sim H_c\cdot T_{\gamma,0}/T_c \sim T_c/M_p \cdot
T_{\gamma,0} \sim 10^{-8} \mbox{ Hz} \]
The absolute maximum amplitude one can expect is $h\sim a/a_0 \sim
10^{-12}$.

We calculated the gravitational wave background normalized to the one
for low frequencies for the standard cosmological evolution
\cite{Schettler:2010dp}. We find that a step in the gravitational wave
background around $\nu \sim 10^{-8}$~Hz is clearly seen resulting from
the QCD phase transition which is about $0.7$. The results are rather
insensitive to the parameterizations used from lattice gauge
calculations and the details of the phase transition. There appears
some more pronounced oscillations for the MIT bag model
parameterization due to the sharper drop in energy density compared to
the lattice data.

The spectrum of gravitational waves for a little inflation will be
drastically different, as the amplitudes are exponentially suppressed
during inflation: $h \sim 1/a \sim \exp (H\cdot t) $. The
gravitational wave background will drop as $\nu^{-4}$ so that the high
frequency part will be a factor $10^{12}$ smaller compared to the one
for standard cosmology. 

Gravitational waves can be presently measured with ground based
interferometers as LIGO, GEO600, TAMA, and VIRGO, which, however, are
mostly sensitive to frequency of about 100--1000 Hz. The future
space-based interferometer LISA will explore the gravitational wave
band around a frequency of $10^{-5}$~Hz. Interestingly, gravitational
waves can be also detected by combining the timing signal for several
millisecond pulsars. First limits for the gravitational wave
background have been already set by Parkes Pulsar Timing Array
\cite{Jenet:2006sv}. The predicted step frequency in the amplitude is
close to highest sensitivity of pulsar timing. In the future the
Square Kilometer Array SKA will increase the sensitivity in this
frequency range by an order of magnitude by measuring hundreds of
millisecond pulsars. The Big Bang Observer BBO, a planned NASA
mission, is aimed at measuring the gravitational wave background from
inflation in a frequency band between the ones of LISA and LIGO (see
e.g.\ \cite{Corbin:2005ny}). In the little inflation scenario, the
gravitational wave background can be detected by looking at the
polarization pattern at the CMB but would be unmeasurably small for
the frequency scales probed by BBO.

Gravitational waves can be generated from bubble collisions at the end
of the little inflation period in addition to the gravitational wave
background.  The gravitational wave amplitude scales then as $h(\nu)
\propto \nu^{-1/2}$ for $\nu < H$ which is just white noise, while it
scales as $h(\nu) \propto \nu^{-2\dots -1}$ for the higher frequencies
within the Hubble horizon $\nu > H$ depending on whether multi-bubble
collisions are taken into account or not
\cite{Kamionkowski:1993fg,Huber:2008hg,Caprini:2010xv}. For the
flatter spectrum, the produced gravitational waves could be measured
with LISA \cite{Schettler:2010dp}.


\section{Summary}


We have explored a cosmological scenario where the early universe
passes through a first order QCD phase transition. For large initial
net baryon numbers so that $\mu/T\sim\cal{O}(1)$ there might be a
first order phase transition line in the QCD phase diagram as
suggested by effective models of QCD. The universe is trapped in a
metastable false vacuum state generating a little inflation with about
seven e-folds.  This little inflationary period would generate
potentially observable signals.

\begin{itemize}

\item The power spectrum of large-scale structure is modified up to
  mass scales of $M \sim 10^9 M_\odot$ (without QCD inflation only
  mass scales up to the horizon mass $\sim 10^{-9}M_\odot$ can be
  affected).

\item The cold dark matter density is diluted by a factor $10^{-9}$ so
  that a reduced WIMP annihilation cross section is needed as
  $\Omega_{\rm CDM} \sim \sigma_{\rm weak}/\sigma_{\rm ann}$ with
  implications for the WIMP searches at the LHC.

\item The first order phase transition generates seeds of
  (extra)galactic magnetic fields by the collisions of charged bubbles
  which would be a viable scenario within the standard model again
  (see \cite{Caprini:2009pr}).

\item The change of the scale factor will modify the gravitational
  wave background by suppressing it for frequencies above about
  $10^{-8}$~Hz. Colliding bubbles and turbulence can generate additional
  gravitational waves which can be observable with pulsar timing and
  eventually with LISA. 

\end{itemize}


\section*{Acknowledgements}

This work is supported by BMBF under grant FKZ 06HD9127, by DFG within
the framework of the excellence initiative through the Heidelberg
Graduate School of Fundamental Physics, the International Max Planck
Research School for Precision Tests of Fundamental Symmetries
(IMPRS-PTFS), the Gesellschaft f\"ur Schwerionenforschung GSI
Darmstadt, the Helmholtz Graduate School for Heavy-Ion Research
(HGS-HIRe), the Graduate Program for Hadron and Ion Research (GP-HIR), and
the Helmholtz Alliance Program of the Helmholtz Association contract
HA-216 ``Extremes of Density and Temperature: Cosmic Matter in the
Laboratory''.


\bibliographystyle{utphys}
\bibliography{newbiblio}

\end{document}